# Autonomous Heavy-Duty Mobile Machinery: A Multidisciplinary Collaborative Challenge

Tyrone Machado
Department of System Development
Bosch Rexroth AG
Elchingen, Germany
tyrone.machado@tuni.fi

David Fassbender
Department of System Development
Bosch Rexroth AG
Elchingen, Germany
david.fassbender@tuni.fi

Abdolreza Taheri
Faculty of Engineering and Natural Sciences
Tampere University
Tampere, Finland
reza.taheri@tuni.fi

Daniel Eriksson
R&D Wheel Loader- Emerging Technologies
Liebherr-Werk Bischofshofen GmbH
Bischofshofen, Austria
daniel.eriksson@liebherr.com

Himanshu Gupta
School of Science and Technology
Örebro University
Örebro, Sweden
himanshu.gupta@oru.se

Amirmasoud Molaei
Institute of Vehicle System Technology
Karlsruhe Institute of Technology
Karlsruhe, Germany
amirmasoud.molaei@partner.kit.edu

Paolo Forte
School of Science and Technology
Örebro University
Örebro, Sweden
paolo.forte@oru.se

Prashant Kumar Rai
Faculty of Engineering and Natural Sciences
Tampere University
Tampere, Finland
prashant.rai@tuni.fi

Reza Ghabcheloo
Faculty of Engineering and Natural Sciences
Tampere University
Tampere, Finland
reza.ghabcheloo@tuni.fi

Saku Mäkinen
Faculty of Management and Business
Tampere University
Tampere, Finland
saku.makinen@tuni.fi

Achim J. Lilienthal
School of Science and Technology
Örebro University
Örebro, Sweden
achim.lilienthal@oru.se

Henrik Andreasson
School of Science and Technology
Örebro University
Örebro, Sweden
henrik.andreasson@oru.se

Marcus Geimer
Institute of Vehicle System Technology
Karlsruhe Institute of Technology
Karlsruhe, Germany
marcus.geimer@kit.edu

*Abstract*—Heavy-duty mobile machines (HDMMs) are a wide range of machinery used in diverse and critical application areas which are currently facing several issues like skilled labor shortages, poor safety records, and harsh work environments. Consequently, efforts are underway to increase automation in HDMMs for increased productivity and safety, eventually transitioning to operator-less autonomous HDMMs to address skilled labor shortages. However, HDMMs are complex machines requiring continuous physical and cognitive inputs from human operators. Thus, developing autonomous HDMMs is a huge challenge, with current research and developments fragmented into several independent research domains. Furthermore, autonomous HDMM technologies are a stack of several technologies requiring a convergence of diverse competencies from the different domains. Through this study, we provide an overview of the HDMM industry and use the bounded rationality concept to propose multidisciplinary collaborations for new developments in autonomous HDMMs. Furthermore, we apply the transaction cost economics framework to highlight the conceptual challenges and implications of these collaborations. Therefore, we bring together several domains of the HDMM industry to introduce autonomous HDMMs as a general and unified approach. The collaborative challenges and potentials are mapped out between the following topics: mechanical systems, AI methods, software systems, sensors, connectivity, simulations and process optimization, business cases, organization theories, and finally, regulatory frameworks. In doing so, we highlight the need for new and multidisciplinary perspectives that should be considered by academic and industrial practitioners working on the development and deployment of autonomous HDMMs.

I. INTRODUCTION

Heavy-duty mobile machines (HDMMs) or mobile machinery are machines used in diverse application areas like agriculture, earthmoving, forestry, etc., which currently face problems of skilled labor shortage, poor safety records, and harsh work environments. To tackle these problems, the HDMM industry is developing automated and autonomous HDMMs. Since the concepts of automation and autonomy are used interchangeably within academic and societal discourse [1], we make a distinction between automation and autonomy. *Automation* is the execution of a function by a machine, in predefined scenarios, with limited decision-making capability, of a task previously performed by a human [1]-[3]. *Autonomy* is a state of high automation wherein the machine senses, perceives, and understands its environment to execute a function in various scenarios using its own decision-making capability [1], [3], [4]. Thus, autonomous HDMMs operate

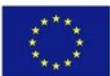
This project has received funding from the European Union's Horizon 2020 research and innovation programme under the Marie Skłodowska-Curie grant agreement No 858101.

without active human intervention. Since automation and autonomy focus on either humans or machines, the concept of augmentation [5] is also important. *Augmentation* is the collaboration of humans and machines together, to achieve a work task, wherein the machine augments human capabilities and vice versa [5].

Although autonomous haulers in the HDMM industry have been operational for decades, the research maturity of autonomous HDMMs is lacking [4]. Research literature addressing autonomous HDMMs is scant while literature addressing general HDMMs is fragmented into several domains. However, research on autonomous driving is quite mature, with current industry trends demonstrating increasing collaborations between automobile manufacturers and suppliers of autonomous driving technologies from unrelated business domains. While HDMMs share similarities with the driving functions of automobiles, HDMMs are available in various shapes, sizes, functions, and other features, thereby exhibiting tailored characteristics, even for machines of the same type [6]. Due to such machine diversity, autonomous machines require a complex stack of several integrated technologies rather than a single technology [7]. Thus, a one-size-fits-all solution for autonomous HDMMs is unlikely.

Considering rapid technological change, environmental pressures [6], [8], and severe labor shortages exasperated by the COVID-19 pandemic [9], interest in autonomous HDMMs has been renewed. However, research on autonomous HDMMs lacks economic investigations and economic models [10]. Furthermore, there is a misconception that autonomous HDMMs/technologies are a sub-field of artificial intelligence (AI), comprising of predominantly embedded systems or software engineers [7], computer scientists, and roboticists [5] researching AI methods to rationalize existing work processes [5]. Thus, inadequately designed education causes talent shortages [7], while insights about AI and autonomous technologies from management scholars are currently limited [5], thereby impacting the research and development (R&D) of autonomous machines [7]. Additionally, since technology entrepreneurship research predominantly uses the biotechnology and semiconductor industry setting, little is known about entrepreneurial activities in other novel areas [11], such as autonomous HDMMs. Autonomous HDMMs can operate in privately controlled environments, thus providing flexibility for the practical implementation of autonomous solutions [12]. Addressing such broad societal, economical, and technological challenges related to autonomous HDMMs require multidisciplinary collaborations [8], due to the complexity of the machines, their processes, and working environments [6]-[7]. Thus, platform-based multidisciplinary collaborations may become important for autonomous HDMM developments, as also evident from a multidisciplinary collaborative platform initiative for autonomous HDMMs in Finland [13]. Such collaborations can be explained using the resource-based view (RBV) and the dynamic capabilities framework, wherein firms adapt and reconfigure their existing resources or acquire new resources to build and sustain their competitive advantage and core competencies [8] in rapidly changing environments [14], [15].

Therefore, due to the convergence of several technical domains required for the realization of autonomous HDMMs, we use the concept of bounded rationality [16] to envision a multidisciplinary collaborative approach for autonomous HDMMs. Here, collaboration is defined as a durable relationship that unites separate technical domains into a new structure, with a commitment to a commonly defined mission (autonomous HDMMs) [17]. Furthermore, the industrial context affects collaboration potentials, thereby making it difficult to extrapolate findings of individual firm-level collaborations from one industry into another industry [18]. Thus, this study aims to introduce autonomous HDMMs as a unified literature stream. Furthermore, this study examines the relations and interactions between different domains of the HDMM industry and provides a conceptual [19] overview of the collaborative efforts required for future autonomous HDMMs. The research questions are formulated as follows: What is the current state-of-the-art in the HDMM industry? How will autonomous HDMM developments foster collaborations in the HDMM industry? What are the implications and challenges of these collaborations from the transaction cost economics (TCE) perspective?

The remainder of the paper is structured as follows. Section II briefly describes the methodology used. Section III describes the theoretical foundations forming the basis of the discussions. Section IV and V introduce the different domains and the current state-of-the-art in the HDMM industry. Section VI outlines the multidisciplinary collaborations and challenges. Section VII presents a discussion on the implications of the multidisciplinary challenges. Finally, section VIII concludes the paper.

II. METHODOLOGY

Research on autonomous machines should consist of at least four major elements of R&D: sensing, perception and planning, client and cloud computing, and mechanical control [7]. Thus, the state-of-the-art in HDMMs and the collaboration potentials in autonomous HDMMs were conceptualized using extant academic literature on mobile machinery and brainstorming sessions among eight researchers in the MORE project [20], working on eight research topics covering the four R&D elements [7]. Accordingly, seven researchers in the MORE project covered at least one of the R&D elements, while one researcher covered the topic of business cases. Moreover, since the MORE project is a consortium of industry and academia, all the eight researchers were/will be affiliated with at least two other industrial actors from the HDMM industry, thereby having practical industrial insights. Each researcher noted their current requirements for HDMMs based on prior literature reviews and industrial discussions. Later, the researchers discussed each requirement individually and identified the common elements between each other's research.

These common elements formed the basis for the future multidisciplinary collaborative challenges mapped out in section VI. Thus, the brainstorming sessions covered the major aspects of collaboration based on current developments. Once the specifications, requirements, and collaborative challenges were mapped out, the implications of these collaborations were analyzed conceptually using the TCE framework [21], focusing predominantly on bounded rationality, opportunism, asset specificity, and transaction costs.

## III. Theoretical Background

TCE deals with exchange relationships related to a particular transaction, i.e., contracting costs [21]. TCE is founded on the assumptions of firstly, bounded rationality, wherein human rationality is cognitively limited with further time constraints, thereby being oblivious to all the possible future scenarios [21]. Thus, all contracts are imperfect; The second assumption is opportunism, wherein every actor who subscribes to an exchange relationship with another actor will try to behave opportunistically with "guile" [21] and exploit vulnerabilities in the exchange. Contracts can be used to mitigate such opportunistic risks. Taken together, this means that every transaction has a cost associated with it known as transaction costs [8], [19], [21], which are affected by transaction attributes such as asset specificity, uncertainty, and frequency [21]. These transaction costs ultimately provide a framework for organizations to draw their boundaries using appropriate governance mechanisms [17], [19], [21].

With globalization, rapid technological change, and new sustainability initiatives, organizations may rely on multidisciplinary collaborations to broaden their capabilities and core competencies [8]. Such collaborations may bring about a change in the boundaries of the firm due to the need for new transactions outside the firm. Typically, inter-organizational collaborations are initiated through formal contracts and collaboration agreements for sense-making and the development of shared goals among the collaborators [22]. However, such agreements may inadequately capture all the conditions for effective collaboration [17], due to bounded rationality and the focus of TCE on individual transactions [21], [23]. Moreover, among the TCE assumptions, opportunism is the least understood [19]. This is important because the nature of collaboration varies across cultures [8], while the effectiveness of contractual safeguards in mitigating the risks of opportunism is dependent on the cultural contexts of the relationship too [19], [24]. This relates to how resources are exchanged among different actors in collaborative networks, with several factors such as government policies, harmonized standards, and geographically relevant economics playing an important role [11] in exchange relationships, which ultimately affect the transaction costs.

Multiproduct firms, for example, the HDMM industry, have close relationships with several actors within the industry who possess co-specialized and/or specific assets. With a rapidly changing market economy, HDMM firms need to make quick strategic decisions to ramp up capabilities [6], and the frequency of firm renewal increases [11], [14]. While TCE places a high focus on asset specificity [21], firms in several HDMM application areas have traditionally maintained stable relationships with other firms [25], and thus, the frequency of transactions is equally important. Moreover, the high-paced, high-tech, innovation economy also brings in high uncertainty which creates additional contractual problems [26]. Furthermore, new information and communication technologies (ICT) can reduce certain transaction costs such as search costs [27]. Thus, TCE alone is insufficient, and complementary theories such as RBV and dynamic capabilities [28], [29] are necessary for a better understanding of the implications of multidisciplinary collaborations in autonomous HDMMs, where considerations must be given to the multiproduct trajectories and technology convergence perspectives [29] of the HDMM industry.

## IV. Domains in Heavy-duty Mobile Machines

Every HDMM performs at least one of the following two core functions: Firstly, manipulation operations using the implements and end-effector(s) of the machine to change the shape, form, size, and/or location of an external material; Secondly, driving/navigation around a worksite. Using these core functions of HDMMs and the structure of the MORE project [20], we propose three primary domains.

### A. Process Domain

Here, HDMMs are considered as part of a broader system such as a worksite. Thus, an HDMM depends on other machines and upstream as well as downstream processes. Since several machines cooperate to reach a common goal, they are known as a multi-agent system. Thus, the efficiency and productivity of an individual HDMM affects the efficiency and productivity of the process itself [4], [20].

### B. Machine Hardware Domain

Here, a single HDMM is the unit of analysis wherein the different mechanical, hydraulic, electric, or hybrid components, and system architectures are considered. Combustion or electric motors, gearboxes, pumps, valves, and cylinders are some of the key hardware components used to move the HDMM and its implements.

### C. Machine Automation Domain

Here, a single HDMM along with its surrounding environment is considered. The focus is on intelligent perception, world modeling, decision making, and automated control in HDMM operations [20]. This domain leverages the use of sensors, datasets, software algorithms, and AI to optimize the core functions of an HDMM and is further divided into two subdomains: *Machine perception* wherein three-dimensional perception, either by the human operator, sensors, or both, is a requirement to understand the HDMM environment; *Machine control* which focuses on the control of the implements and end-effectors of HDMMs. Here, optimization-based algorithms are used to assist the human operator and/or to automatically control the machine.

## V. State-of-the-art in Heavy-duty Mobile Machines

Each domain of HDMMs has some current industry and research specific requirements which are outlined further.

### A. Requirements for Process Domain

*1) Manufacturer-supplier-customer relationship:* The value proposition of HDMMs influences the choice of HDMM for a customer, depending on worksite tasks, resource and capital availability, market conditions, and competitive bidding. Such requirements determine the size and type of the HDMM used. The HDMM industry is a low volume, high product cost industry. Thus, the system development and integration costs are very important [6], wherein it is vital for original equipment manufacturers (OEMs) to differentiate themselves from the competitors. OEMs and their suppliers specialize in different HDMM domains and may possess both, generic and co-specialized assets. Thus, OEM-supplier relationships such as the level of vertical integration [21], along with the exploitation of valuable, rare, inimitable, and non-substitutable (VRIN) resources [15], have a significant effect on the industrial relationships and competitive edge.

*2) Work performance evaluation:* Currently, the skills of human operators highly influence the performance of HDMMs [30], thereby requiring performance monitoring systems and computer-aided management tools. Additionally, due to changing environmental and machine load conditions, real-time performance evaluation, condition monitoring, and fault detection methods are desired, which require high computational power of the embedded hardware, depending on the computational complexity of the evaluation method. Furthermore, HDMM operations have repetitive work cycles, thus, small improvements in operation cycle time or fuel efficiency bring about significant improvements in the overall performance of HDMMs [30].

*3) Planning and workflow optimization:* As work progresses, HDMM operations need to be replanned in real-time, which requires planning strategies. Furthermore, HDMMs interact with external materials having specific properties and behaviors, in different weather conditions. Thus, a specific HDMM or cooperation of different HDMMs may be required to manipulate the materials. Additionally, environmental conditions on worksites can be very dynamic, thereby affecting the HDMM's ability to execute the planned operations. Thus, maintaining an updated representation of a worksite is crucial.

### B. Requirements for Machine Hardware Domain

Traditional human operated HDMMs are as large as possible to increase the productivity per operator. Moreover, large HDMMs are usually powered by diesel, require large actuator stroke lengths, high force, as well as high power capabilities, making hydraulics the state-of-the-art technology. However, recent trends in automation and tightening emission regulations have shifted the focus towards hybrid and electrified powertrains [6]. For human operated HDMMs, predefined work hours per shift with short breaks from work are standard. Since electric batteries require long charging times, utilizing electric batteries instead of diesel engines as the energy source is challenging because an electrified fleet of HDMMs requires additional charging infrastructure, which may further require asset-specific investments and new transactional relationships [23]. Furthermore, certain powertrain and actuator technologies induce noise, vibration, and harshness (NVH), thereby requiring more comfortable operator cabins, which require significant design efforts and coordination between engineers and cabin designers.

### C. Requirements for Machine Automation Domain

The human operator is responsible for the control and navigation of the HDMM and may be provided with automated assistance functions, for augmentation [5]. However, since existing HDMM systems and control architectures are not particularly designed for computerized systems [4], the sensors have an ad-hoc placement. This is important because sensors must cover a 360° field of view for effective collision-free navigation and manipulation. Furthermore, data generated from the different sensors needs to be collected and fused together so that control algorithms can optimize the core HDMM functions.

*1) Machine perception:* Environment perception for autonomous navigation involves solving some of the following problems: mapping the environment, localization of the HDMM in the environment, obstacle detection for collision avoidance, traversable path detection, and HDMM self-awareness. The environment perception system consists of sensors like cameras, lidar, and radar. Cameras and lidar provide detailed information about the environment. Cameras are susceptible to varying illumination and weather conditions, while lidar works in low illumination but provides noisy data in adverse weather conditions. Radar is more resilient against adverse weather conditions but is inherently difficult to use. Furthermore, sensors such as inertial measurement units, wheel encoders, and global navigation satellite systems, are also important in HDMMs. These sensors are used along with perception sensors for the precise localization of the HDMM in the environmental map.

*2) Machine control:* Currently, the human operator perceives the environment and controls the HDMM implements accordingly. Material properties (e.g., granular size and density), work tasks, and modes of operation are consequential for the automation of control systems. Besides the perception system, the machine incorporates various sensors to keep track of the forces, velocities, and positions of the implements during operations. Human capabilities pose two limitations in machine control: Firstly, while human operators attempt maneuvers in the shortest time possible (by training instructions or intuition), this optimization is not a quantified solution. Conversely, state-of-the-art algorithms can plan the operations using mathematically quantified optimizations, providing more efficient operations. Secondly, operators as boundedly rational human beings have insufficient reaction time and processing capabilities [5], [16]. The commands provided by human operators using levers, handles, or joysticks are passed on to low-level controllers that command the actuators of the machine. These controllers stabilize the system while compensating for the unwanted NVH factors. Such features in the control systems augment [5] the human operator ease-of-use without contributing to full automation or energy efficiency. Thus, future autonomous control will be less constrained by factors addressing human comfort.

## VI. MULTIDISCIPLINARY COLLABORATION AND CHALLENGES

The progress from no automation to autonomy in HDMMs will proceed in a stepwise manner with intermediate augmentation [5] steps like operator assistance and teleoperation systems [4]. Accordingly, several collaboration potentials and challenges are described further.

### A. Role of the Human Operator

While the human operator will be vital in the future, the dynamics in a worksite and the operator's tasks will change due to machine augmentation of human capabilities [5], thereby having several macroeconomic and microeconomic implications [10]. Thus, the role of the human will change from being an operator of one HDMM to a coordinator of several automated/autonomous HDMMs. In the absence of a human operator, condition monitoring and fault detection methods become crucial. Finally, task planning/coordination will also be automated in the future, wherein the human would transform into a supervisor of several autonomous HDMMs. Here, the main challenge pertains to the transaction costs related to training and educating new personnel [7] for various autonomous HDMM operations.

## B. Process Planning and Optimization

The machine automation domain uses material and environmental representations to model and simulate the interactions and effects of external materials on HDMM implements. Thus, the process domain and the machine automation domain require close collaborations to develop and use standardized models/representations of the environment and to potentially increase the efficiency and productivity of a worksite. By measuring the performance of different autonomous HDMMs, it is possible to identify a bottleneck HDMM, and subsequently plan and coordinate the operations on a worksite for optimal performance, by optimizing one of the core functions of the HDMM. Thus, even though the productivity of a particular HDMM will be reduced, the productivity of the whole process will remain unaffected.

## C. Mechanical and System Architectures:

With novel autonomous worksites, smaller HDMMs become more attractive [10] because multiple, smaller, driverless machines can be used flexibly [9]. Thus, the average stroke lengths, maximum forces, and actuator power requirements will decrease. Consequently, actuator concepts such as electro-mechanical linear spindle actuators should be considered too. Smaller machines have lower power and energy requirements per machine, which may improve the feasibility of electric batteries as primary energy sources. The benefits of electric machines range from low NVH operations to lower maintenance requirements. However, OEMs and their suppliers need to consider the lifecycle management of electric powertrains because electrical components such as high voltage systems are not widely available while embedded software systems become obsolete faster than traditional technologies [6].

Optimized process planning would also enable the electrification of larger HDMMs. Thus, expensive fast chargers could be used more continuously, by planning the charging breaks of multiple HDMMs sequentially rather than concurrently. Moreover, cabinless HDMMs require lenient NVH considerations, which increases the feasibility of different powertrain technologies as well as higher actuator speeds and acceleration limits. Machine control concepts have a significant impact on the powertrain design too. Typically, HDMMs use multiple actuators which allow the end-effectors to reach certain positions with an indefinitely large set of trajectories. For example, if centralized valve-controlled actuators are used, sequential actuator movements are more efficient than simultaneous movements. If an autonomous controller considered such changes, the efficiency of powertrains could be significantly improved, both reliably and consistently. Furthermore, a traditional assumption for the design of HDMMs is that usually no more than three actuation axes are controlled simultaneously. With automated control, machines with many more axes can be controlled in new and efficient ways [31]. Thus, the number of actuators required for a single HDMM might rise significantly and design parameters such as costs per actuator become more influential, favoring specific actuator concepts [32].

## D. Sensors

The placement of the different sensors in HDMMs depends on the type and size of the machine. Autonomous HDMMs would require a redesign of the machine suited for computerized control systems [4]. Environment perception could be improved by using multiple types and configurations of sensors, which in practice, are heavily limited/influenced by the cost of sensors. For example, three-dimensional (3D) lidars are relatively expensive sensors and provide very dense 3D information about the environments, but cannot be used in adverse weather conditions (like snow, rain, fog) or dust, which is a fairly common working condition for HDMMs. Environment perception may also be affected by the vibrations from HDMMs interfering with sensor outputs. Thus, although a cabinless machine may have lenient NVH requirements, NVH must be considered in novel system architectures for autonomous HDMMs.

## E. Software Systems and Control Algorithms

New and robust perception solutions, as well as control methods, utilize AI methods such as machine learning and neural networks, which require specialized high-performance hardware, for example, graphical processing units. Furthermore, machine control and perception strategies use software algorithms, datasets, and data fusion, which also place high requirements on the computational hardware [6]. Additionally, methods such as imitation learning (IL) are used to construct surrogate models for control of an autonomous HDMM, wherein a human operator demonstrates a task, which is subsequently "imitated" by the model [33]. The human performance may differ between operators and thus, work performance evaluation models are valuable to demonstrate ideal working methods for IL. High computational hardware and embedded hardware requirements also drive up the costs required for additional infrastructure such as data storage systems [6]. A review of autonomous trucks indicated that 85% of the additional costs required to make trucks autonomous come from highly specialized software [34]. Similarly, a study on construction machinery demonstrated the use of servitization by an OEM to reduce implementation costs for the end-user by using a machine-to-machine cloud connectivity solution [35].

## F. Simulation and Validation

Due to bounded rationality [16], software algorithms are limited by the cognitive abilities and experience level of their respective developers [5], thereby requiring validation. Codifying and rationalizing existing processes is complex and depends on several domain-specific experts' tacit knowledge [5], which requires multidisciplinary perspectives to resolve the ambiguities arising due to bounded rationality [5]. Such algorithms and models are validated and optimized using virtual or semi-virtual simulations of the real world, thereby requiring proper and accurate representation of datasets from the other HDMM domains.

## G. Connectivity

Communication architecture plays an important role in autonomous HDMMs. Advancements in ICT reduce transaction costs related to searching information [27], paperwork, and bureaucracy [36]. For autonomous HDMMs, connectivity along with the internet of things (IoT) becomes important since it's required not only for communication and planning between HDMMs but also for tele-remote operations. Moreover, high initial investments in IoT technologies and the lack of interoperability among IoT providers hinder the adoption of IoT-based solutions for end-users with smaller operations [36]. Several machine-type communications (MTC) and communication architectures such as 3G, 4G, and 5G, were analyzed in [37]. A standardized connectivity infrastructure would enable more efficient real-time online planning, site updates, coordination of HDMMs, condition monitoring, fault detection of machines, and cloud computing.

## VII. Discussion

Autonomous HDMMs have many implications and collaborative potentials. R&D of perception sensors within autonomous driving is very mature. With perception sensors such as 3D lidars being expensive and the HDMM industry being a low volume industry [6], the HDMM industry must leverage such R&D from autonomous driving. Furthermore, autonomous technologies are a stack of several technologies [7], while several HDMM application areas are privately owned, devoid of public scrutiny [12], with certain HDMM OEMs and their suppliers having a strong presence in both industries. Thus, close collaborations between the two industries allow diverse testing and data collection scenarios for future sensor developments.

Augmentation [5] of human operators in HDMMs will affect the existing relationships between OEMs, suppliers, and customers. Hydraulics as the state-of-the-art technology in HDMMs creates path dependencies for future innovations, wherein the HDMM industry pursues efficiency rather than radical innovations to meet market requirements [29]. However, the shift towards electric or hybrid powertrains [9] requires a new lifecycle analysis [6] of transaction costs related to switching technologies. Thus, existing asset-specific investments which yielded knowledge transfer and a steady flow of rents between the transactional partners [23], would now be disrupted due to newer technology and knowledge requirements.

In autonomous HDMMs, software and algorithms are more likely to provide a competitive edge compared to hardware alone in traditional HDMMs. Thus, OEMs and suppliers would require new VRIN resources [15] to maintain a competitive advantage. Thus, the boundaries of the firms may have to change not only with reference to TCE contractual issues [26] but also by developing dynamic capabilities [18], [26]. Furthermore, the effect of contracts on tacit knowledge acquisition is limited [22], and considering bounded rationality [16], it is important for organizations to stimulate different perspectives to coding by including data scientists, business managers, engineers [5], and social scientists to develop the right autonomous/AI capabilities [26], [29]. Such capabilities may be developed by hiring new talent or training existing talent with the right competencies [7], thereby incurring additional training costs. While larger firms can invest in training new students at universities to tackle talent shortages, many firms working on autonomous technologies are early-stage companies with limited resources [7], [38]. Furthermore, the rarity of economic models for autonomous HDMMs [9], high initial investments required [10], and short obsolescence cycles of AI-related technologies [6] may have led to a hype that draws attention towards future R&D funding [6] for technologies related to autonomous HDMMs. Additionally, organizations may pursue automation for short-term cost efficiencies, forcing competitors to follow suit [5]. For example, competitive bidding/tenders for new mining equipment have an "autonomous capability" clause, which automatically excludes HDMM firms not offering any autonomous functionalities.

Considering these hurdles, platform-based ecosystems could be a suitable alternative for future R&D in autonomous HDMMs, wherein a focal firm provides a platform on which ecosystem collaborators can build/develop/test their own technology offerings [39], thereby leading to open innovation breadth and depth, sharing of resources and knowledge, and complementary technological capabilities [18]. The focal firm concentrates on innovations supported by upstream component/service suppliers, while downstream complementors enhance the value of the product/service offering for the customers [39], thereby co-creating value for the ecosystem. Typically, since complementors are considerably smaller than incumbents [40], for example, new technology-based firms (NTBF) [11], [38], they cannot rely on vertical integration to prevent larger incumbent technology-based firms (ITBF) [11] from offering similar products [40]. On the other hand, ITBF move slower due to their larger size, path dependencies, and higher switching costs associated with new technologies [28]. Thus, a collaborative platform-based ecosystem could enable ITBFs such as OEMs and their suppliers from the HDMM industry, and NTBFs such as robotics and AI technology start-ups, to co-create value using ecosystem governance and by leveraging the resources within the ecosystem [39]. Furthermore, platform-owners/incumbents should seek to improve the ecosystem by targeting underperforming complementors rather than focusing on existing product spaces of successful complementors [40], thereby ensuring the sustainability of the ecosystem.

Such collaborative ecosystem governance may be economically more viable, rather than negotiating new contracts, especially since the HDMM industry would be competing with the automotive industry for similar resources, skills, and technologies. However, there are transactions costs associated with identifying, selecting, and screening collaboration partners, especially if the platform-owner wants to ensure knowledge protection [18]. Owners of intellectual property may withhold their strategic assets in the presence of direct competitors [29], which increases the costs for protection activities especially when there is a high diversity of collaboration partners [18]. Conversely, compared to the industry, universities and R&D centers can afford the risks of investing in highly specialized technologies, making them important collaboration partners in ecosystems [18], for example, as evident in the "PEAMS: Platform Economy for Autonomous Mobile Machines Software Development" project initiative [13] for standardized software development.

Some application areas of HDMMs exhibit poor cost performance, where little is known about project transaction costs, even though they account for a significant part of the project [25]. Considering the end-user, autonomous HDMMs may require legally mandated on-site supervision [9], with associated transaction costs, both for the client and the end-user [25]. Moreover, autonomous HDMMs open new possibilities like data analytics, condition monitoring, etc. New business models may be needed to leverage these possibilities. For example, transaction costs related to site visits of clients can be reduced [25] by real-time online machine data monitoring and updates. Concurrently, with the current state of R&D, autonomous HDMMs will be more expensive than traditional HDMMs. Thus, alternative business models such as servitization and solutions-oriented business models may become important in the future. However, a study [28] highlighted that TCE implications and the technological intensity in an industry may reduce a firm's profitability for solutions offerings. Since the HDMM industry has high technological intensity, the implications and viability of new business models for autonomous HDMMs need thorough analyses. Autonomous HDMMs may also lead to the use of AI in the management of worksites. Traditional management theories are very human-centric, revolving around human-agency [5], but machines behave rationally even though humans have a social element to their behavior [5]. Thus, hybrid organizational systems with open rational (intelligent machines) and open natural (human-agency) systems, and differing organizational boundaries, require new managerial perspectives, to account for new assumptions related to human-machine augmentation [5].

An additional challenge in autonomous HDMMs is within the domain of safety and security and has also been addressed as an interdisciplinary challenge for the automotive industry [41]. Safety considerations and compliance with regulations are paramount for the successful deployment of autonomous HDMMs [10]. A lack of harmonized standards related to autonomous HDMMs leads to different geographical regulations, which impact the transaction costs related to compliance, costs related to compulsory on-site supervision of autonomous HDMMs [9], and compliance with new standards [42]. Furthermore, compliance costs for non-harmonized standards are predominantly borne by OEMs, which increases the transaction costs associated with market delays [42]. With the use of AI in autonomous technologies, several nations are introducing new regulations for AI in machines. Such regulatory changes impact the HDMM industry in terms of the existing contracts and additional compliance costs. Furthermore, data is very important for AI and IoT. With the lack of proper regulations and interoperability standards [36], cyber security and trust are important factors for autonomous HDMM adoption [36]. Cyber security in autonomous HDMMs has its own challenges and addresses the topics of data confidentiality, integrity, and availability, which need to be addressed right from the early design stages [12].

## VIII. CONCLUSION

The current state-of-the-art in the HDMM industry reflects a fragmentation of HDMM R&D into several technical domains, with OEMs and suppliers specializing in different domains. With the drive towards autonomous HDMMs being led by augmentation, and autonomous technologies being a stack of several technologies, new multidisciplinary perspectives are required which cross the boundaries of the different domains. However, collaborations across such boundaries require considerations on the transaction costs, especially with platform-based ecosystems, where firm boundaries may get blurred. This poses several challenges in the effective collaboration required for the R&D and deployment of autonomous HDMMs. The theoretical implication of this study is that future researchers working on autonomous HDMMs could use the interactions mapped out in this study, to understand the implications of their own research on other domains, thereby fostering closer cooperation/collaborations between academia and industry, to improve the handling of the practical challenges in autonomous HDMMs. Furthermore, this study consolidates several keywords related to HDMMs, thereby introducing autonomous HDMM as a unified literature stream. The managerial implications highlight the need for new resources, competencies, and knowledge from a TCE perspective, as well as fresh managerial perspectives to deal with AI-enabled autonomous HDMM deployments.